\documentclass[a4paper,11pt]{article}
\pdfoutput=1 

\usepackage{jinstpub} 
\usepackage{graphicx} 
\usepackage{upgreek}
\usepackage{wrapfig,subfig}

\title{\boldmath Hardware Demonstrator of a Compact First-Level Muon Track Trigger for Future Hadron Collider Experiments }

\author[a,1]{D. Cieri,\note{Corresponding author.}}
\author[a]{S. Abovyan,}
\author[a]{V. Danielyan,}
\author[a]{M. Fras,}
\author[a]{P. Gadow,}
\author[a]{O. Kortner,}
\author[a]{S. Kortner,}
\author[a]{H. Kroha,}
\author[a]{F. M\"uller,}
\author[a]{S. Nowak,}
\author[a]{P. Richter}
\author[a]{and K. Schmidt-Sommerfeld}

\affiliation[a]{Max-Planck-Institut f\"ur Physik\\
F\"ohringer Ring 6, 80805 Munich, Germany}

\emailAdd{davide.cieri@cern.ch}

\abstract{Single muon triggers are crucial for the physics programmes at hadron collider experiments. To be sensitive to electroweak processes, single muon triggers with transverse momentum thresholds down to 20 GeV and dimuon triggers with even lower thresholds are required.

In order to keep the rates of these triggers at an acceptable level these triggers have to be highly selective, i.e. they must have small accidental trigger rates and sharp trigger turn-on curves. The muon systems of the LHC experiments and experiments at future colliders like FCC-hh will use two muon chamber systems for the muon trigger, fast trigger chambers like RPCs with coarse spatial resolution and much slower precision chambers like drift-tube chambers with high spatial resolution.

The data of the trigger chambers are used to identify the bunch crossing in which the muon
was created and for a rough momentum measurement while the precise measurements of the muon
trajectory by the precision chambers are ideal for an accurate muon momentum measurement.

A compact muon track finding algorithm is presented, where muon track candidates are
reconstructed using a binning algorithm based on a 1D Hough Transform. The algorithm has been
designed and implemented on a System-On-Chip device. A hardware demonstration using Xilinx
Evaluation boards ZC706 has been set-up to prove the concept.

The system has demonstrated the feasibility to reconstruct muon tracks with a good angular
resolution, whilst satisfying latency constraints. The demonstrated track-reconstruction system, the chosen architecture, the achievements to date and future options for such a system will be discussed.}


\begin{document}
\maketitle
\section{Introduction}
The exceptional results obtained by the Large Hadron Collider (LHC) \cite{LHC} at CERN are pushing towards the development and the design of next-generation hadron colliders. CERN has already planned to upgrade the LHC to the High-Luminosity LHC in order to reach a luminosity up to one order of magnitude higher than the nominal design \cite{HLLHC}. In addition to that, studies have begun for a conceptual design of a Future Circular Collider (FCC) with a centre-of-mass energy $\sqrt{s}=100$\,TeV \cite{FCC}. 

Experiments at future colliders will have to operate in a very harsh radiation environment and with high detector occupancies. The physics programme of these experiments will still be focused on the understanding of the electro-weak symmetry breaking mechanism and on the search of Beyond Standard Model physics signals.

Selective muon triggers will still be crucial to undertake this ambitious physics programme. The foreseen high luminosities will require highly selective first-level (L1) muon triggers in order to obtain the output trigger rate within acceptable rates. This can be done by improving the spatial resolution of the triggering system, resulting in a drastically sharpened turn-on curve of the L1 trigger efficiency with respect to the muon transverse momentum $p_T$. 

\section{First-Level Muon Triggers at future colliders}
A typical muon detector system at high energy hadron collider physics experiment is composed of two kinds of gas detectors. Fast gaseous detectors, e.g. Resistive Plate Chambers (RPCs) or Thin Gap Chambers (TGCs), which have an excellent time resolution (better than 20\,ns), ideal to identify the bunch crossing in which the muons have been produced; and high-resolution detectors, like cylindrical Drift Tubes (DTs) that provide more precise measurements of the muon positions. 

Three possible arrangements for the muon chambers have been taken into account in this paper.

\begin{figure}[hbpt]
\centering
\includegraphics[width=10cm]{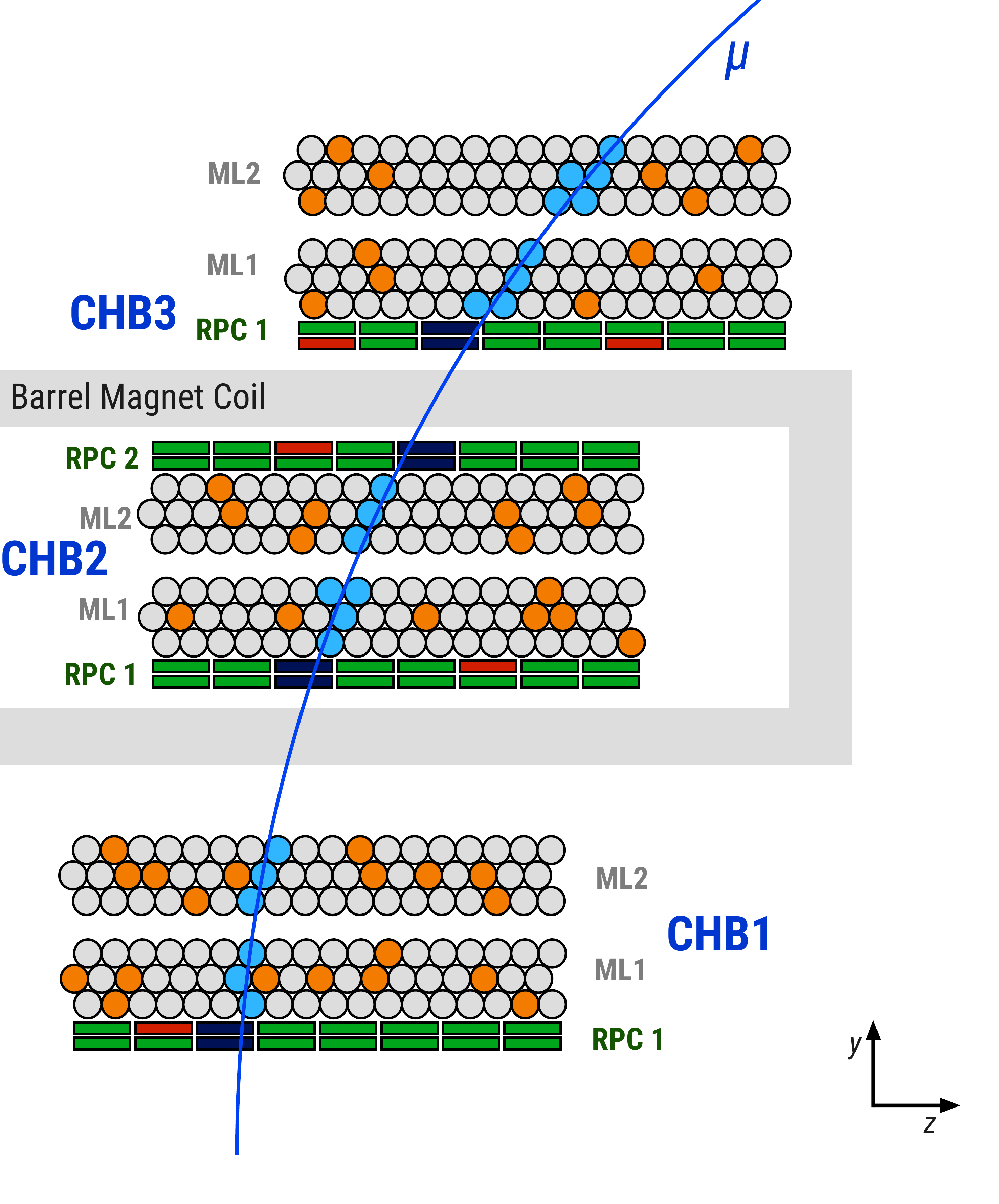}
\caption{An ATLAS barrel-like muon spectrometer. The system is composed of three layers of DTs and RPCs immersed in a magnetic field.}
\label{atlas-barrel}
\end{figure}

\begin{figure}[hbpt]
\centering
\includegraphics[width=10cm]{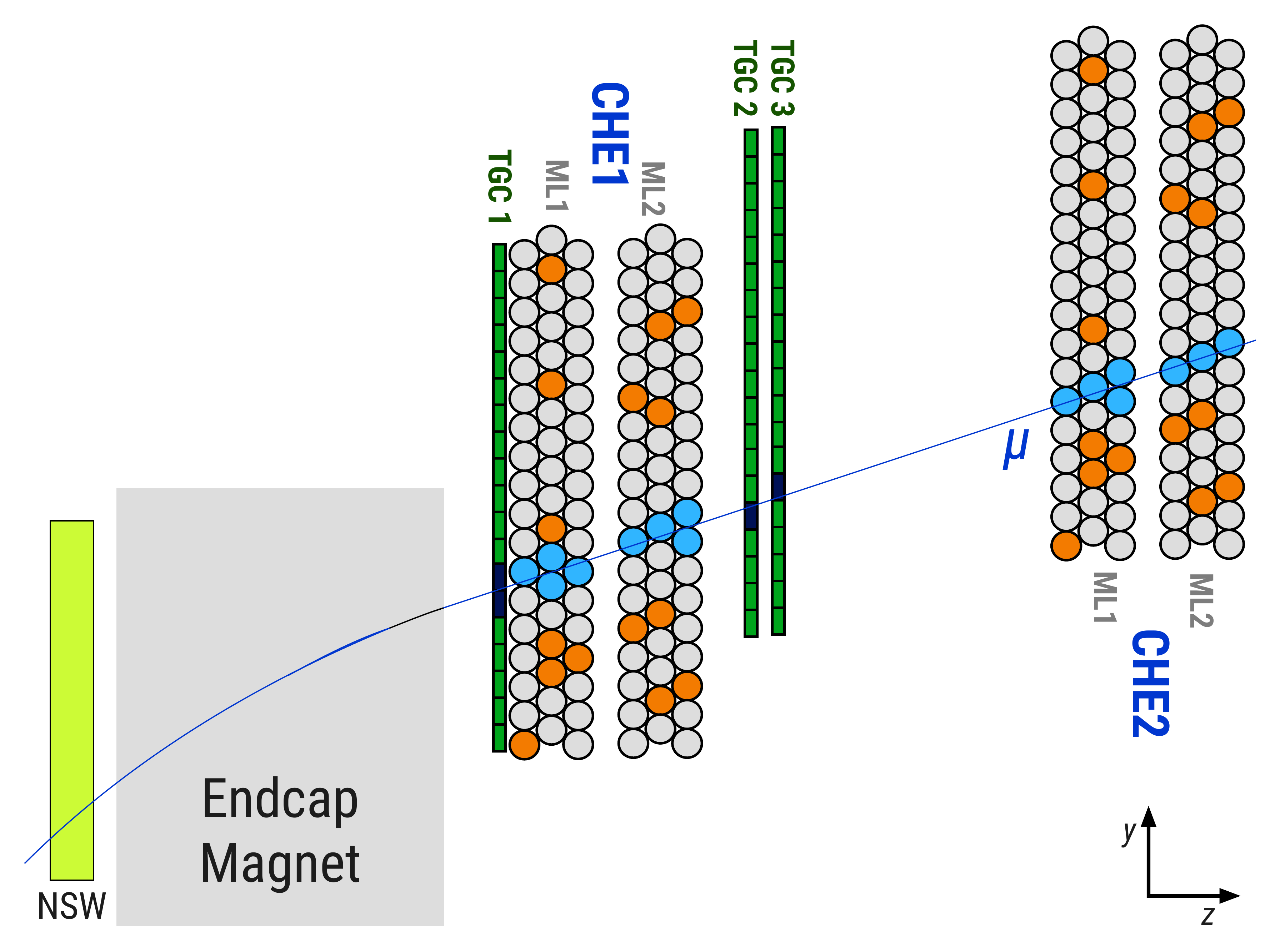}
\caption{An ATLAS endcap-like muon spectrometer. The system is composed of two layers of DTs and TGCs located in a field-free area. The addition of an extra chamber before the magnet (e.g. the New Small Wheel in ATLAS) allows the computation of the muon transverse momentum.}
\label{atlas-endcap}
\end{figure}

\begin{figure}[hbpt]
\centering
\includegraphics[width=10cm]{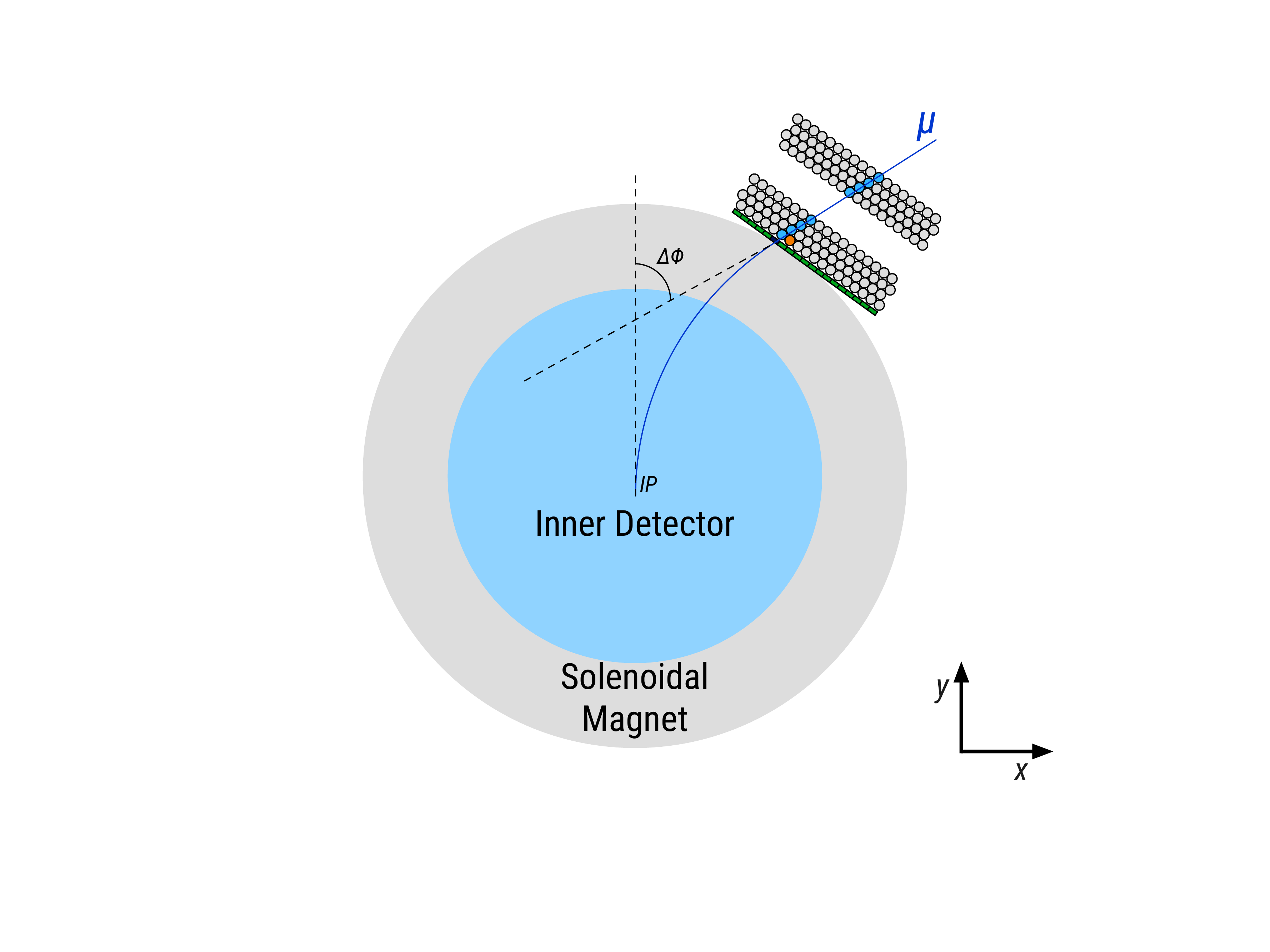}
\caption{An FCC-like muon spectrometer. The system is composed of DT and RPC devices located after the magnet in a field free area. The muon transverse momentum can be computed by extrapolating the reconstructed muon segment to the interaction point (IP).}
\label{fcc-muon}
\end{figure}

\begin{itemize}
\item \textbf{ATLAS Muon Barrel (Figure \ref{atlas-barrel}).} The system is composed of three layers of muon chambers immersed in a magnetic field and disposed at distance of the order of few meters. Each chamber (CHB1, CHB2, CHB3) is composed of two multi-layers (ML1, ML2) of DTs, surrounded by fast RPC devices. Reconstructing the track segments in the different chambers, it is possible to calculate the deviation angle or the sagitta of the muon track and, eventually, to compute the muon transverse momentum.
\item \textbf{ATLAS Muon End-cap (Figure \ref{atlas-endcap}).} In this case, two layers of muon chambers are installed in a magnetic field free area, where muons have a straight line trajectory. The monitored DT detectors are composed of two multi-layers and are surrounded by moderately fast TGC devices, capable to work in high-rate regions, like the ATLAS endcaps. The transverse momentum can be computed by adding a third muon chamber before the magnet (e.g. New Small Wheel in ATLAS) and by computing the deviation of the position measurement of the new chamber from the extrapolated line connecting the two DT chambers.
\item \textbf{FCC/CMS Muon System (Figure \ref{fcc-muon}).} A likely possibility for the FCC experiment consists of adopting a CMS-like approach, utilising a single DT chamber composed of two multi-layers of four tube layers separated by 1.5\,m for first-level trigger purposes. The DT detectors will be surrounded by RPC devices, and it will be located after the super-conductive magnet. The muon transverse momentum will be then calculated measuring the deviation between the reconstructed muon segment and the impact parameter (IP).
\end{itemize}

\begin{figure}[hbpt]
\centering
\includegraphics[width=12cm]{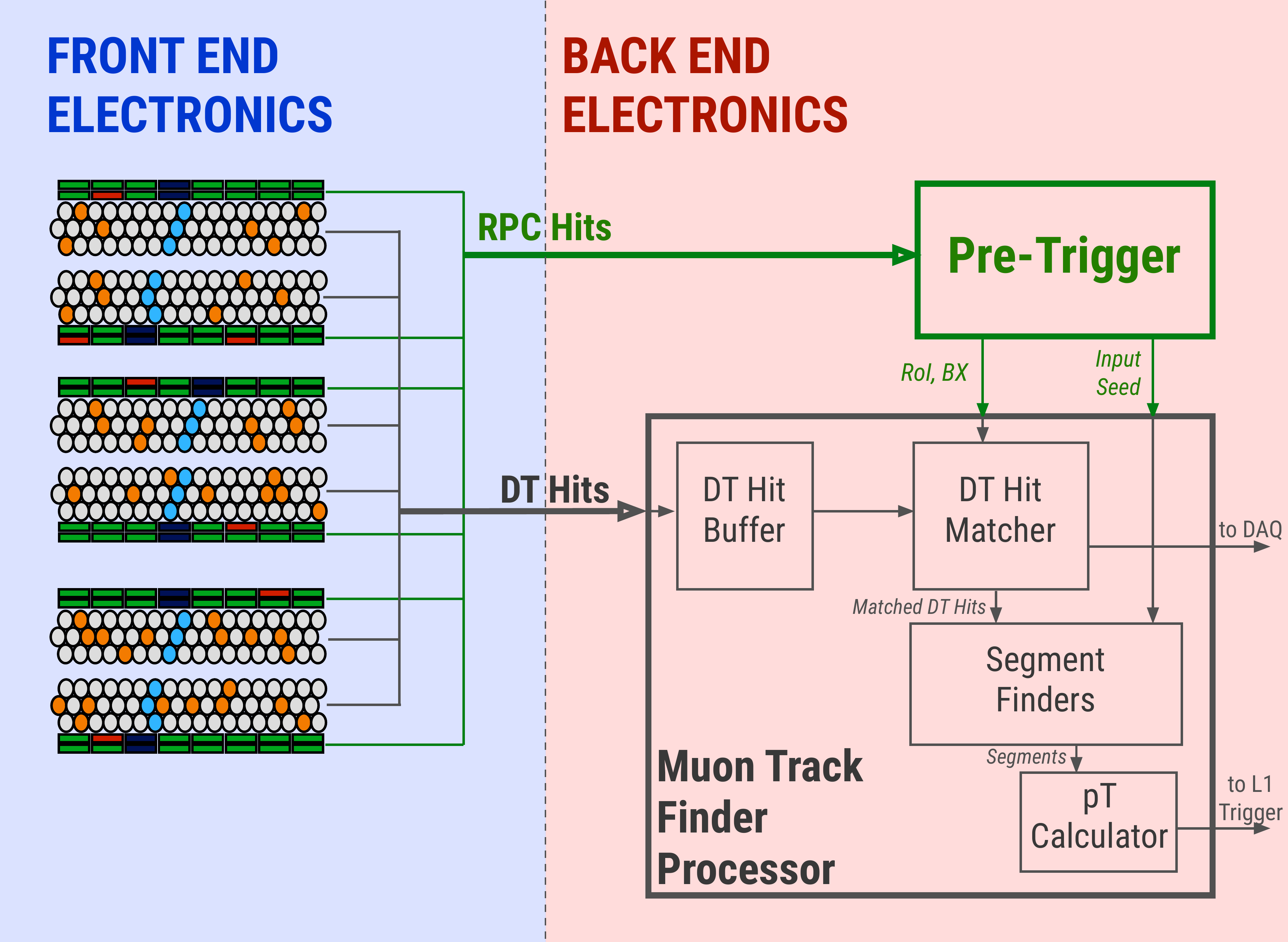}
\caption{Diagram of a possible readout chain for a muon trigger sector. RPCs hits are transmitted first to a \emph{pre-trigger} processor, which identify the RoI and BX and computes a coarse estimation of the segment parameters. Meanwhile, DT hits are transmitted to the Muon Track Finder Processors, where they are matched with the pre-trigger windows and then used to perform a more precise segment reconstruction. The muon transverse momentum is then computed and the result is forward to the L1 Global Trigger for a final decision. }
\label{dataflow}
\end{figure}

An L1 Muon Trigger system at future colliders will make use of both fast and precise chambers to identify the segments and select high-$p_T$ muons. Typical thresholds for single muon and dimuon triggers are 20 and 10\,GeV. The Muon Trigger could be divided into two sub-systems: a \emph{pre-trigger}, which processes data from the fast chambers and has the role to identify region-of-interests (RoIs) compatible with a high-$pT$ muon candidate and with the bunch crossing (BX) under investigation. Once the RoIs have been identified, they are transmitted to the downstream Muon Track Finder Processor (MTFP). Figure \ref{dataflow} shows a schematic data-flow of the L1 Muon Trigger system.

All the hits that have been fired in the DTs, together with the information coming from the pre-trigger, will be input to the MTFPs. The first step is then to select the DT hits in each chamber that are compatible with the identified BX and RoI. These hits are then used to perform a more accurate pattern recognition, which keeps only those hits that belong to the actual segment and uses them to compute the required segment parameters. Once the segments have been found, it is possible to calculate the muon transverse momentum and make an L1 trigger decision.

There are several possibilities to perform the segment reconstruction in the MTFPs. The goal of the segment reconstruction algorithm is to identify the segments from high-$p_T$ muon track in the DT chambers with high efficiency and good resolution. The algorithm must also be fast, to operate within the typically short first-level trigger latency, and be hardware implementable. Finally, the designed algorithm must be lightweight in terms of resource utilisation, in order to keep the cost of the final system low.

\section{A Compact L1 Muon Track Finder Processor}

A lightweight, fast, efficient and cost-effective algorithm is presented in this article. The algorithm is based on the use of the Hough Transform (HT) \cite{Hough} technique to perform pattern recognition. 




\subsection{Segment Finder with a 1D Hough Transform}
Within a single multi-layer DT chamber, the magnetic field integral is so small that the muon trajectory can be approximated by a straight line,

\begin{equation}
y = mz + b,
\end{equation}
where $(z,y)$ are the local coordinates of the muon in the chamber at each time, and $(m,b)$ are the slope-intercept parameters of the segment. A Hough Transform considers a straight line in terms of its slope-intercept parameter. Each point on the track draws a straight line in the HT space. A crossing of lines in the HT space identifies the $(m,b)$ parameters corresponding to a possible track candidate.

Each fired drift tube in the DT chamber measures the drift radius $r$, corresponding to the distance between the tube anode ($z_t,y_t$) and the muon track,

\begin{equation}
r=\frac{mz_t +b + y_t}{\sqrt{1+m^2}}.
\end{equation}

In addition to the RoI information, the pre-trigger could be used also to compute a first estimate of the segment parameters ($\bar{m},\bar{b}$) that can be used as input seed by the segment finder algorithm. Indeed a 1D HT array can be constructed, where the $m$-axis reduces to the point $m=\bar{m}$. Since the two MLs are supposed to be at a distance of the order of 20 tube diameters, it is reasonable to consider them independents and to have a histogram for each ML.

\begin{figure}[hbpt]
\centering
\includegraphics[width=7.5cm]{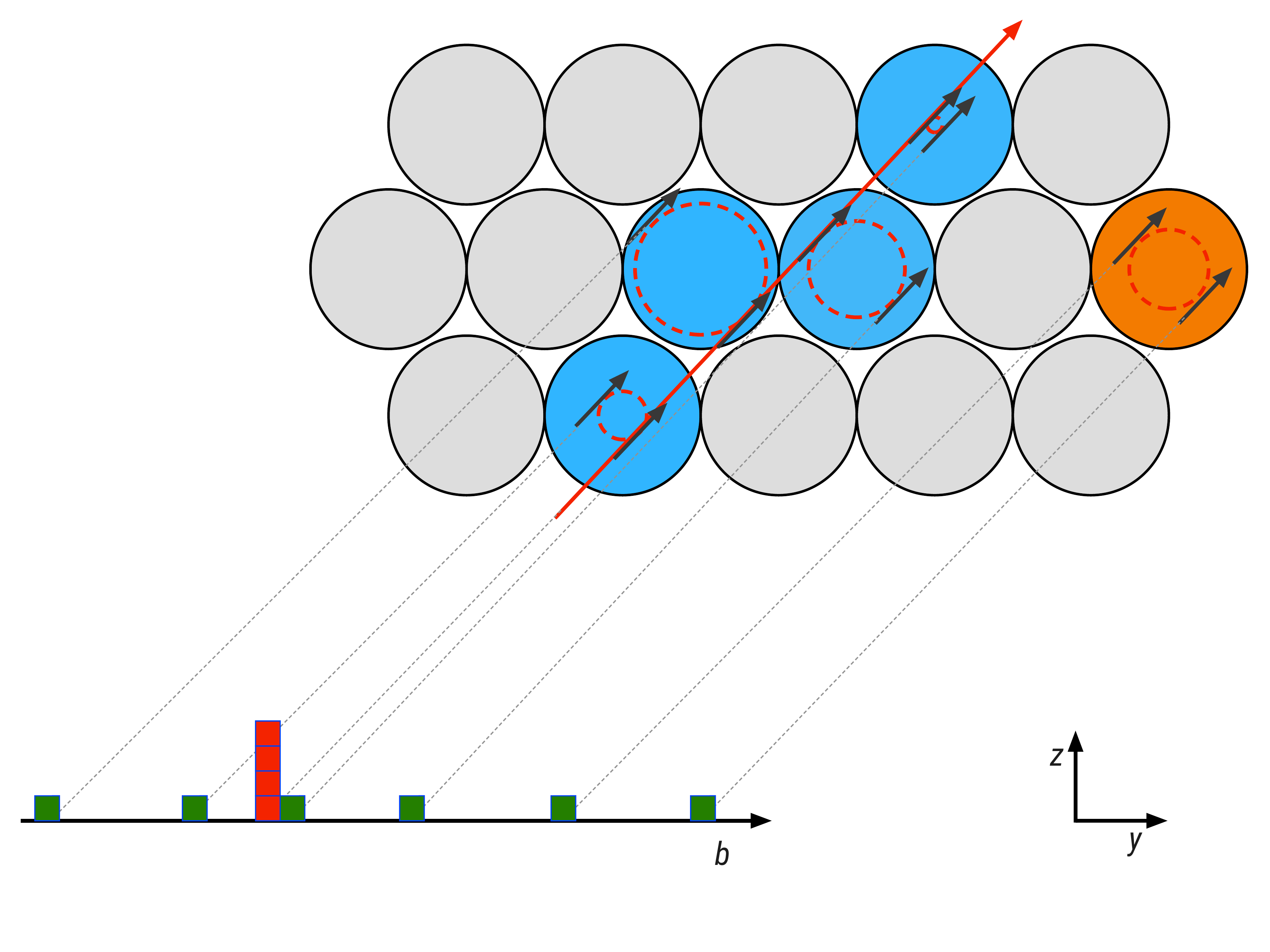}
\caption{Illustration of the 1D Hough Transform method. For each fired tube, two values of $b_\pm$ are computed because of the two-fold ambiguity of the drift tube. A 1D histogram of $b$ is then filled. A segment candidate is then identified by the bin with the largest hit content. }
\label{ht}
\end{figure}

For each DT hit a value of the intercept $b$ is calculated,

\begin{equation}
b_\pm = \pm \sqrt{1+\bar{m}^2}\cdot r - (\bar{m}z_t-y_t),
\end{equation} 
where the sign's difference is due to the two-fold ambiguity of the drift tube. Each hit fills then the 1D HT histogram with its values of $b_\pm$. A segment candidate is identified if at least one maximum per multi-layer is found, where a maximum shall contain at least two hits. If more than one maximum with different hit content is present, more segment candidate can be generated. A limit on the number of maxima per ML histogram can be defined to reduce bandwidth and resources. Figure \ref{ht} illustrates the HT pattern recognition work-flow for a segment candidate in one ML. 

Hits belonging to a segment candidate are finally extracted and used in the downstream segment fitting stage.

\begin{figure}[hbpt]
\centering
\includegraphics[width=7cm]{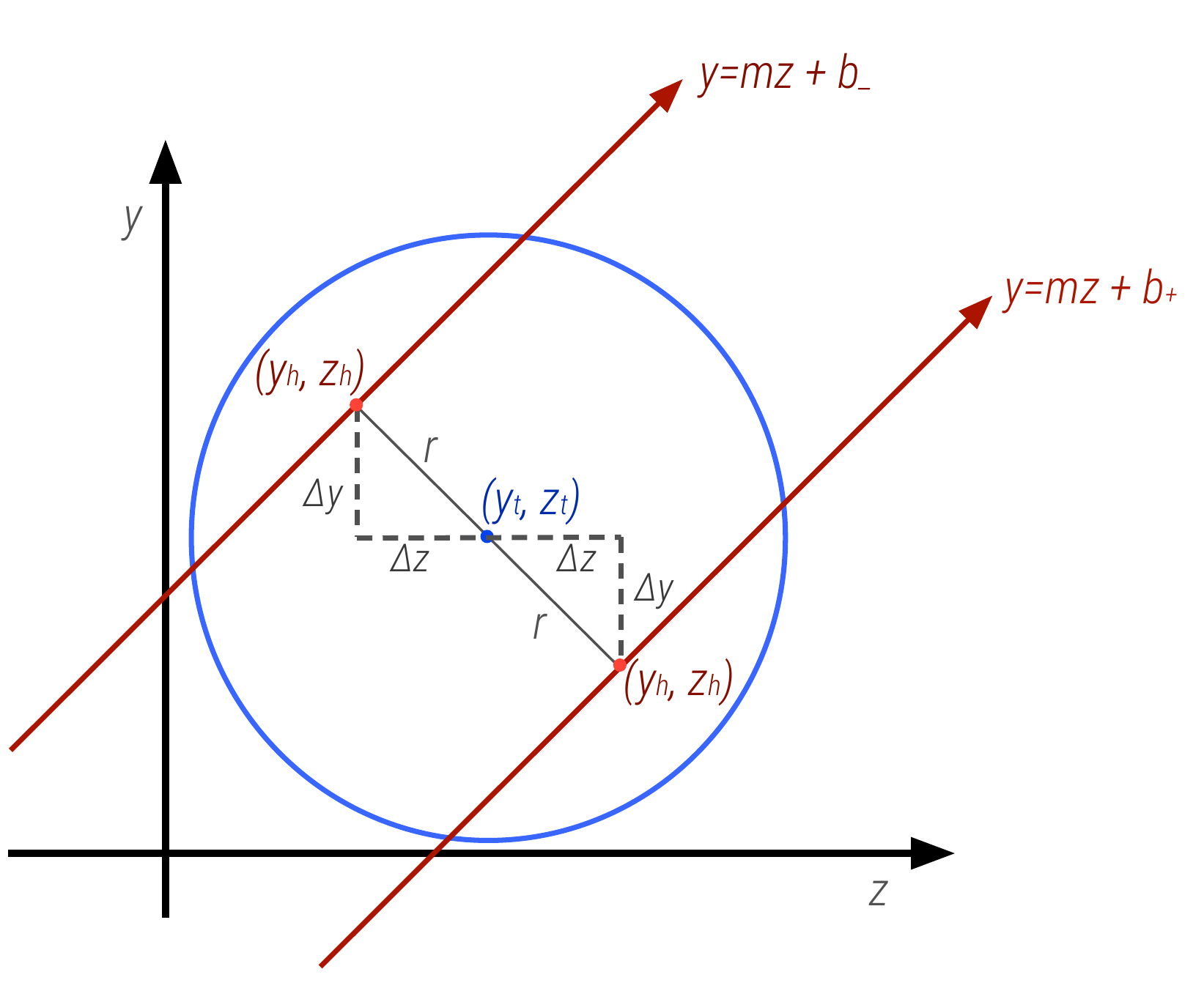}
\caption{Illustration of the hit coordinates and their relation to the tube parameters.}
\label{hit_coord}
\end{figure}

\subsection{Segment Fitter with a Simple Linear Regression Technique}
In the case where there is no magnetic field within a chamber or the field integral is small, the muon trajectory in a chamber can be treated as a straight line. In this case, the segment parameters are calculated by means of a Simple Linear Regression (SLR) fitter. 

Once the segment candidate hits have been identified, it is possible to compute the exact hit positions on the segment,

\begin{equation}
y_h = y_t \pm \Delta y,\qquad z_h = z_t \mp \Delta z,
\end{equation}
\noindent where the signs of $\Delta y,\; \Delta z$ depend on the which solution of $b_\pm$ is compatible with the investigated segment candidate. Figure \ref{hit_coord} shows a scheme of the utilised coordinates for a single tube.

$\Delta y$ and $\Delta z$ can be calculated by means of simple trigonometric rules that express them in terms of $\bar{m}$,

\begin{equation}
\Delta y  = \frac{r}{\sqrt{1+\bar{m}^2}}, \qquad
\Delta z  = r \cdot \frac{\bar{m}}{\sqrt{1+\bar{m}^2}}.
\end{equation}

Once the hit coordinate have been calculated, it is finally possible to compute the segment fit parameters using the standard linear regression formulas,

\begin{align}
\hat{b} &= \frac{(\sum_h y_h)(\sum_h z_h^2) - (\sum_h z_h y_h)(\sum_h z_h)}{n_\text{Hits} (\sum_h z_h^2)-(\sum_h z_h)^2},\\
\hat{m} &= \frac{n_\text{Hits}(\sum_h z_h y_h)- (\sum_h y_h)(\sum_h z_h)}{n_\text{Hits} (\sum_h z_h^2)-(\sum_h z_h)^2},
\label{fit-param}
\end{align}
\noindent where $n_\text{Hits}$ is the number of hits in the current segment candidate.

In the case that more than one segment has been reconstructed, the algorithm can set-up in different ways. For single muon triggers, it keeps only the segment with the lowest fit $\chi^2$, while for a dimuon trigger it would keep the two segments with the lowest $\chi^2$ values. In principle, a cut on the $\chi^2$ could be also be imposed to reject fake segments.

Segments are reconstructed independently in the three muon chambers and then used for the final $p_T$ computation.

\subsection{Transverse Momentum Calculation}

Different strategies can be used to compute the muon transverse momentum, depending on the number of measured positions in the detector. The transverse momentum of a muon particle with charge $q$ travelling in a homogeneous magnetic field of magnitude $B$ is defined as,

\begin{equation}
p_T = qBR,
\end{equation}
where $R$ is the curvature radius $R$. Different methods can be adopted to measure $R$, depending on the number of available position measurements in the detector. 

\begin{figure}[hbpt]
\centering
\includegraphics[width=10cm]{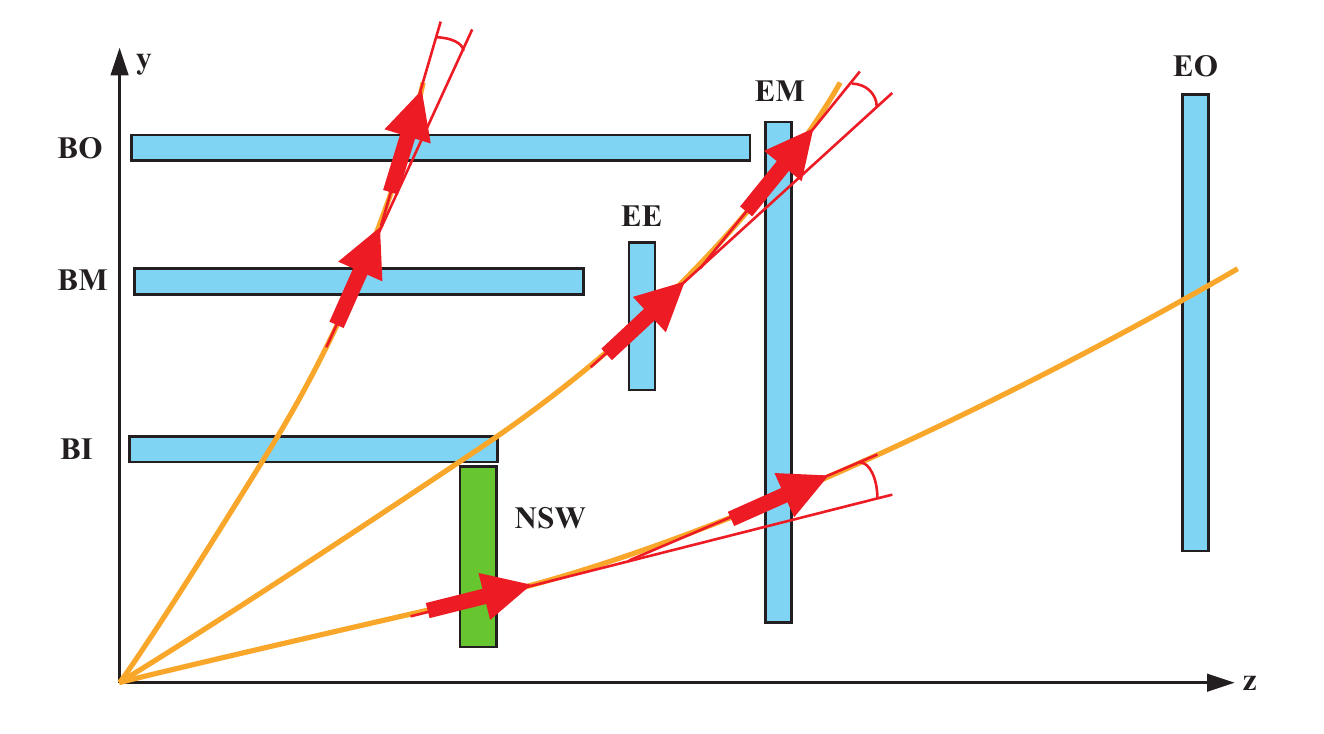}
\caption{Schematic view of the deflection angle measurement in the ATLAS detector \cite{ATLAS_TDAQ}.}
\label{deflection_angle}
\end{figure}

If only two measurements are available, the curvature radius can be calculated by measuring the deflection angle $\Delta\theta$, defined as the polar-angle difference between segments in two chambers. 

\begin{equation}
R = \frac{\Delta y}{2\sin\Delta\theta},
\end{equation}
where $\Delta y$ is the distance along the $y$-axis between the two positions (Figure \ref{deflection_angle}). 
 
\begin{figure}[hbpt]
\centering
\includegraphics[width=10cm]{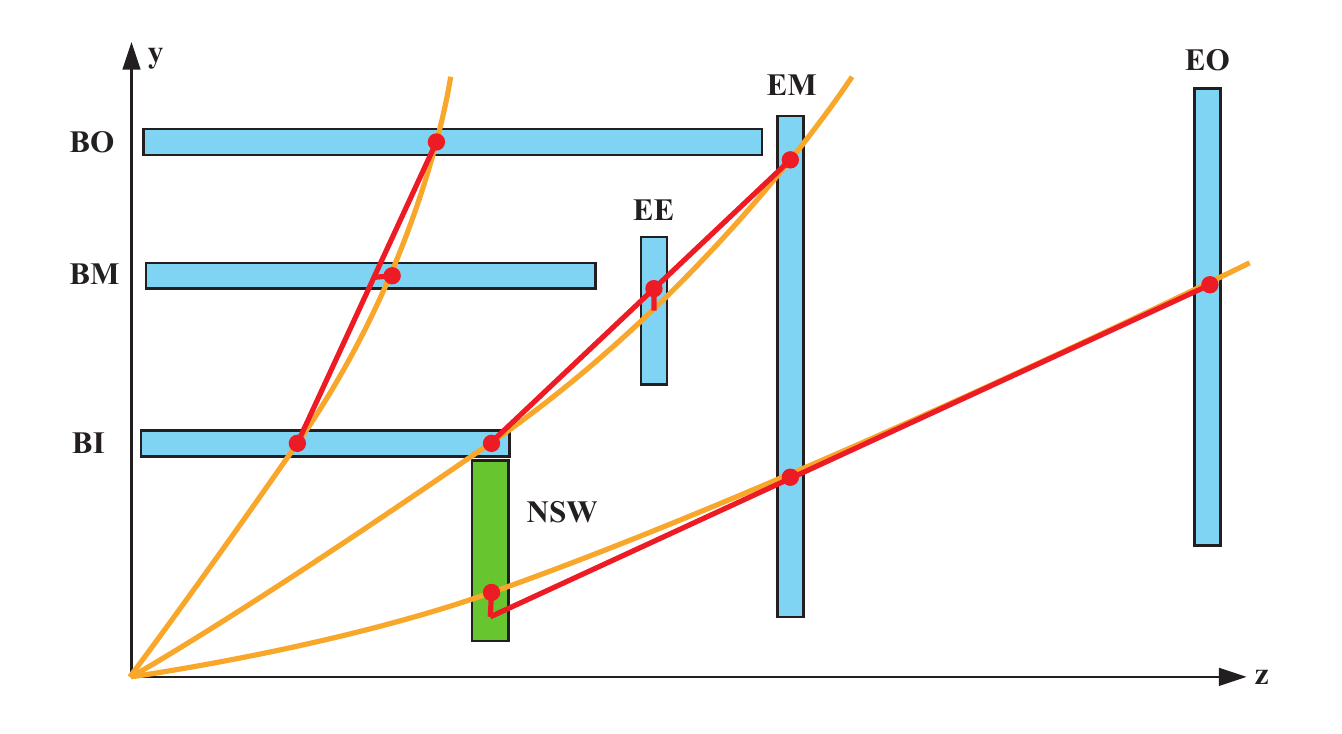}
\caption{Schematic view of the sagitta parameter measurement in the ATLAS muon spectrometer.  \cite{ATLAS_TDAQ}.}
\label{sagitta}
\end{figure}

If segments are available in three chambers, the \emph{sagitta} method \cite{Gadow} can be used. The sagitta $s$ is defined as the distance between the position of one segment from the straight line joining the other two (Figure \ref{sagitta}). For small incident angles, the curvature radius $R$ can be approximated as,

\begin{equation}
R = \frac{8s}{(\Delta y)^2}.
\end{equation}

The previous equations are valid for a system immersed in a homogeneous magnetic field. Unfortunately, this is not the situation for the ATLAS barrel muon system, where the magnetic field presents strong variations. In this case, the detector volume can be divided into small regions and the magnetic field $B(\phi,\eta)$ can be approximated to the second order of the series expansion. The transverse momentum can then be expressed as,

\begin{equation}
p_T = \frac{q\Delta y^2 B(\eta,\phi)}{8s} = S(s) + P(\phi) + E(\eta),
\end{equation}
where $S,P$ and $E$ are simple polynomial functions,

\begin{align}
S(s) &= (1/s - a_0)/a_1,\\
P(\phi) &= \sum_{i=0}^2 p_i \cdot \phi^i,\\
E(\eta) &= \sum_{i=0}^2 e_i \cdot \eta^i,
\end{align}
with $a_i,p_i$ and $e_i$ parameters that can be determined numerically with experimental data.

\section{Hardware Demonstrator}
In order to demonstrate the feasibility of the approach and measure its performance, a full muon reconstruction at hardware trigger level in a single trigger tower (three muon chambers) have been implemented for the System-on-Chip (SoC) Xilinx Zynq device\footnote{Xilinx xc7z045ffg900-2}, mounted on a Xilinx ZC706 evaluation board \cite{Zynq}.

The algorithm has been adapted to the geometry of the Phase II upgrade of the ATLAS muon barrel spectrometer \cite{ATLAS_TDAQ}. So, the current muon trigger system, composed by RPCs and TGCs, is used as a pre-trigger for the MTFP. MDT hits are then matched and used for the finer segment reconstruction \cite{ATLAS_MDT}.

\begin{figure}[hbpt]
\centering
\includegraphics[width=9cm]{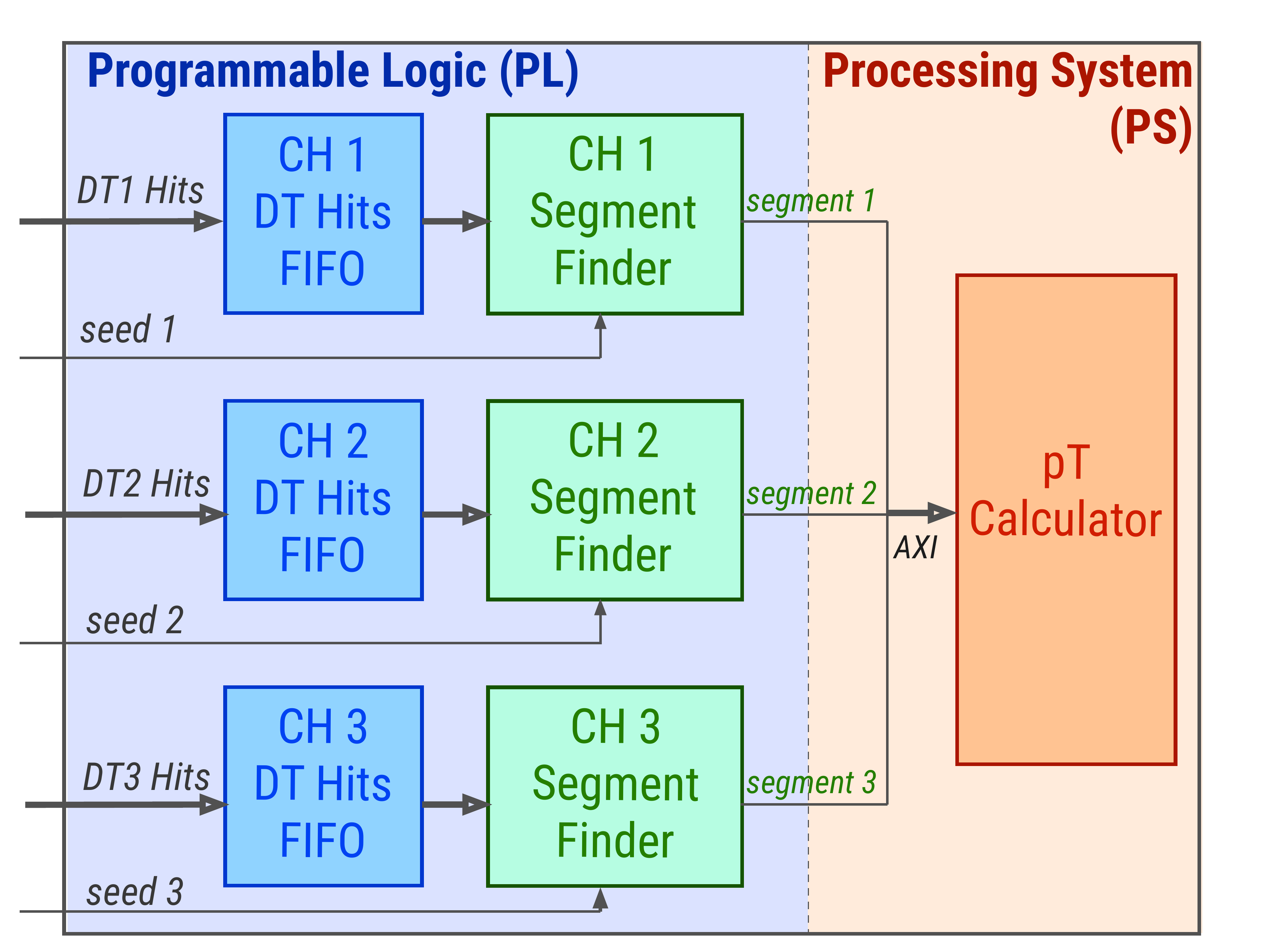}
\caption{Overview of the main components of the MTFP and their location on the Zynq device. DT hits from each chamber are first stored in a FIFO, waiting for the arrival of the pre-trigger seed. Once it comes, hits are transmitted to the segment finder modules, which reconstruct the three segments. Segments are then transmitted to the PS via AXI bus where the final $p_T$ calculation takes place. An alternative approach integrates also the $p_T$ calculation in the PL area.}
\label{fpga}
\end{figure}

Figure~\ref{fpga} shows an overview of the MTFP, illustrating the main logic components and their interconnectivity. The pattern recognition and the segment fitting steps are implemented as dedicated IP-Core into the Programmable Logic (PL) of the device, while two different implementations of transverse momentum computation have been investigated. A first possibility consists of performing the $p_T$ calculation into the Processing System (PS), where it could take advantage of the floating-point unit (FPU) and the media processing engine (NEON) of the ARM processor. An alternative would be integrating it directly on the PL, which eliminates any further latency due to data transmissions between the PL and the PS.

\subsection{Demonstrator Framework}

\begin{figure}[hbpt]
  \centering
  \includegraphics[width=9cm]{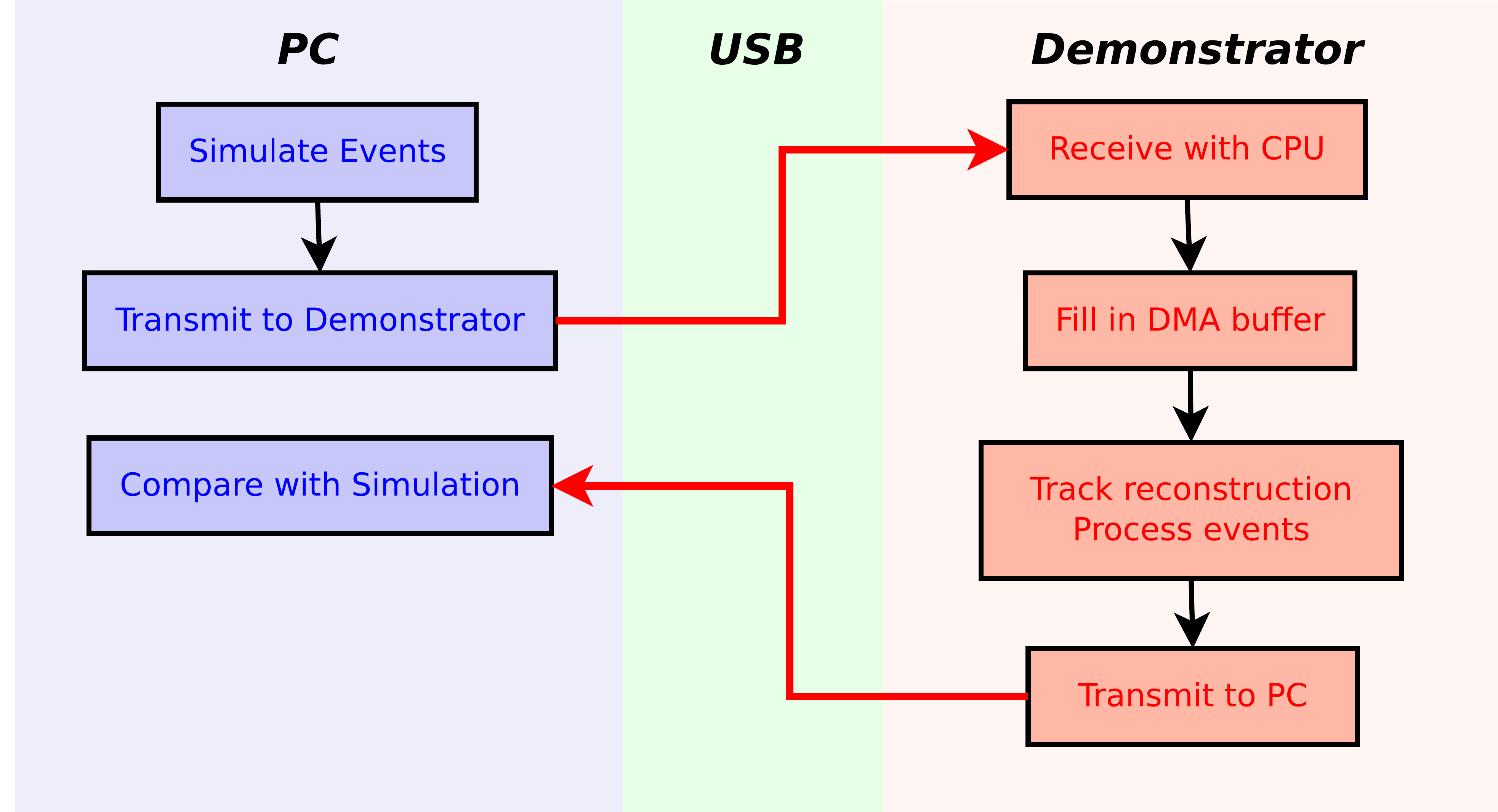}
  \caption{Dataflow of the demonstrator set-up. Simulated data is transferred via USB to the demonstrator where it is processed by the track reconstruction logic. Afterwards, the results are transferred back to the PC where they are compared with simulation.}
  \label{fig::dataflow}
\end{figure}

In order to measure the performance of the implementation under realistic condition and with sufficient statistics, simulated single muon events are used as input data and are transmitted from PC software via USB to the PS and further to the PL by using a DMA IP-Core (Direct Memory Access), which loads the single hit information one-by-one into the trigger IP-Core.
After track reconstruction, the resulting data is returned to the PS and further back to the PC where the results are compared with simulation, see Fig.~\ref{fig::dataflow}.

Instead of transmitting the full dataset via USB, one could consider to store input data in the PS memory and, therefore, minimize implementation effort.
As the PS memory provides a two-level cache, the timing performance of this simple implementation would be sufficiently better than for a realistic implementation, because the input data is permanently stored in the cache, and, therefore, immediately available for processing.

\subsection{Algorithm Implementation in FPGA}
\begin{figure}[hbpt]
\centering
\includegraphics[width=12cm]{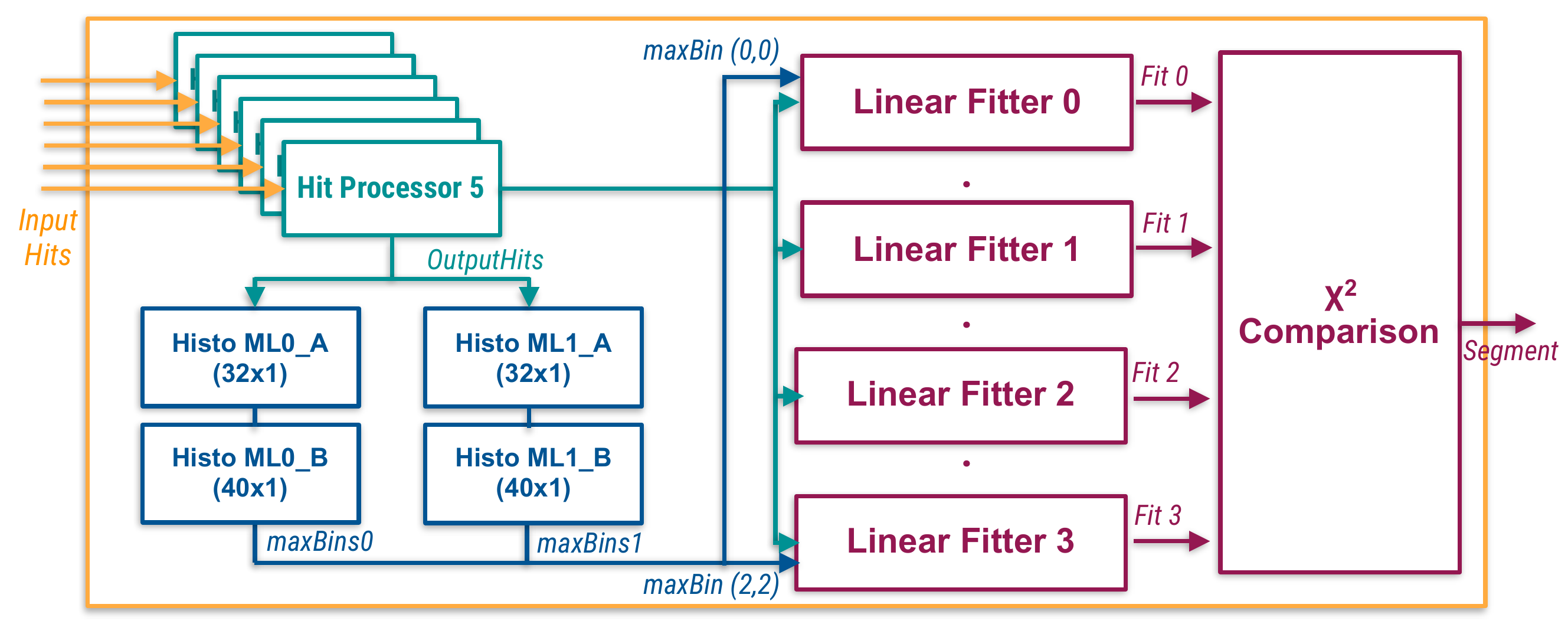}
\caption{Firmware implementation of one segment finder processor. Hits belonging to different tube layers are sent to independent Hit Processor modules for the $b_\pm$ computation. Hits are then transmitted to the two-stage Histogram blocks, which identify the maxima in the two ML. Hits belonging to maxima are then forwarded to the Linear Fitters, for the final segment parameter calculation. A $\chi^2$ comparison is then implemented to select only the segment with the better fit quality, which is eventually selected and transmitted to the $p_T$ calculation unit.}
\label{fw}
\end{figure}

Figure \ref{fw} shows a simplified layout of the implemented segment finder firmware. The RoIs of a future ATLAS or FCC first-level muon trigger are supposed to have a maximum width of $\pm3$ tubes. Assuming that the tubes will be readout per tube layer, with each tube sending one hit per clock, a fixed-length packet of six clock cycles is input to the firmware. The seed information is assumed to arrive before the first hit comes in. 

Hits are directly transmitted to the \emph{Hit Processor} modules, which compute for each hit the values of $b_\pm$ and forward them to the two HT histograms, one per ML, for the pattern recognition stage. 

To achieve acceptable efficiency, it has been found that a 1D HT histogram of 256 bins is necessary. Implementing such a big histogram straight away into the FPGA would require a large use of LUT. On the other hand, the pattern recognition step could be easily divided into two stages.

\begin{enumerate}
\item Hits fill first histograms of 32 wider bins of 7.5\,mm width, covering the entire range of $b_\pm$ (\emph{Histo ML0\_A, Histo ML1\_A}). The maximum in the histograms are then identified and all hits belonging to bins in an interval of $\pm 2$ bins from the maximum are selected and transmitted to the second histogram.
\item Second histograms of 40 bins of 0.93\,mm width, the same granularity of the 256-bins histogram, are then filled with the surviving hits (\emph{Histo ML1\_B, Histo ML1\_B}). This histogram covers a fixed smaller range of $b_\pm$, whose limits change at each time depending on the position of the maximum in the first histogram. In order to avoid bin-boundaries effect, each hit fills also the bins corresponding to values of $(b_\pm \pm 1)$ for a total width of 2.8\,mm (\emph{triple-filling}).
\end{enumerate}

Each bin in the histograms keeps a record of the hits so that possible duplications are avoided. Up to two maxima per histogram are identified, meaning that a maximum of four segment candidates can be produced. Hits belonging to candidates are then transmitted to the \emph{Linear Fitter} instances, which compute the segment fit parameters and the $\chi^2$ values. The Simple Linear Regression fitting stage has been implemented using a chain of DSP blocks. The divisions present in Eq. \ref{fit-param} have been implemented using a 21-bit depth ROM.

The implemented algorithm has been designed for a single muon trigger. For this reason, produced segments are compared in the $\chi^2$ \emph{comparison} module, which selects the segment with the lowest $\chi^2$ and transmits it to the $p_T$ calculation unit.

\subsection{Demonstrator Results}
To validate the implemented algorithm, simulated physics events of single muons were produced with standalone simulation software, including modelling of the ATLAS Phase-II MDT chambers and of the expected background hit rate at HL-LHC. For these studies only the more difficult case of the ATLAS Muon Barrel spectrometer has been considered, where the provided input seed has an average angular resolution of the order of the 10-15\,mrad.

An \emph{emulation} software has been developed, processing the same digitised hit and seed information used as input to the demonstrator to reconstruct segments. Hit and seed data are transmitted to the evaluation board over USB-UART and output segments are read out in the same way. A comparison software has also been created, comparing the results obtained in hardware and emulation. Since the developed \emph{emulator} code is not clock-accurate in the description of the algorithm, small differences are expected to arise.

\begin{figure}[hbtp]
     \centering
     \subfloat[][]{\includegraphics[width=8cm]{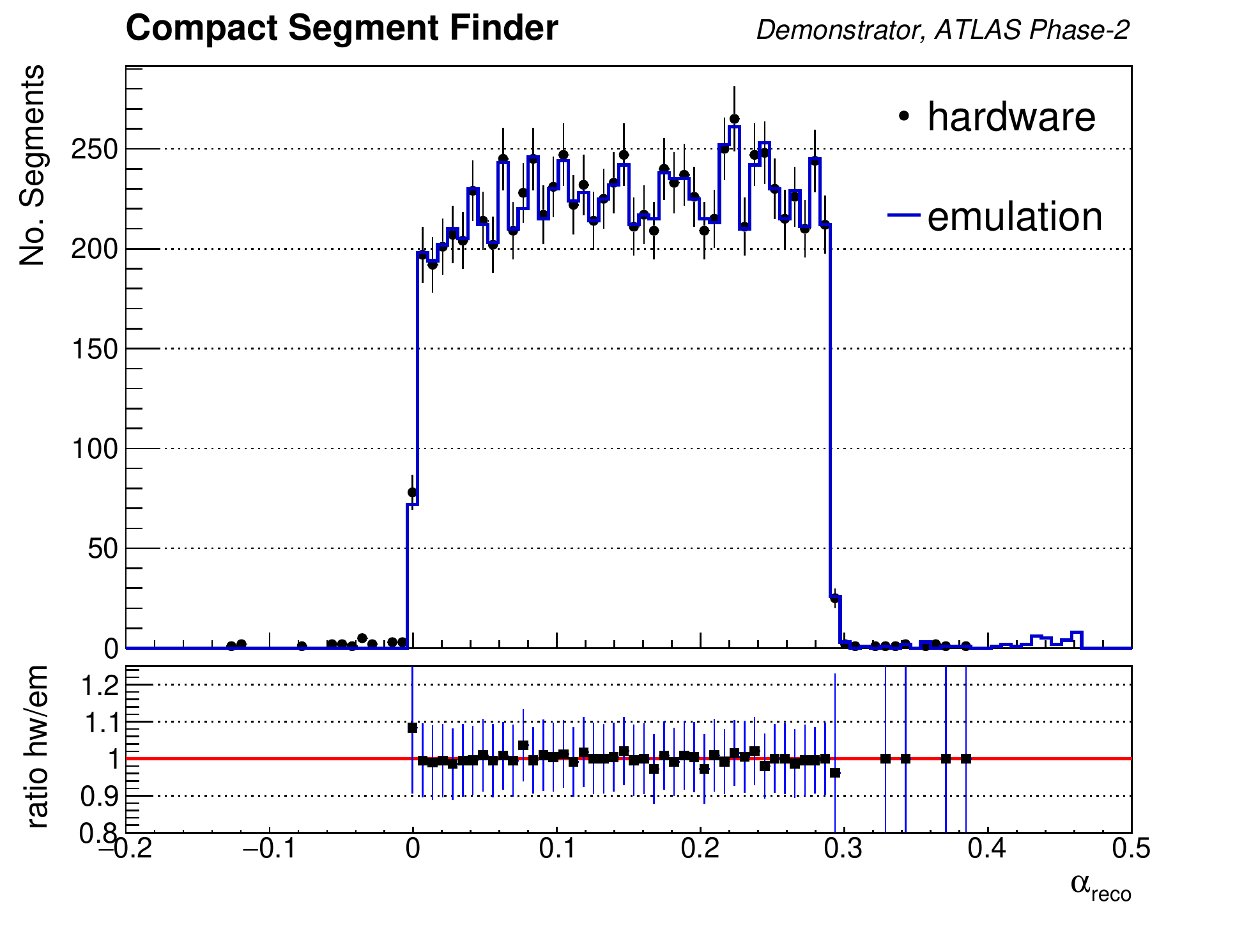}\label{<figure1>}}
     \subfloat[][]{\includegraphics[width=8cm]{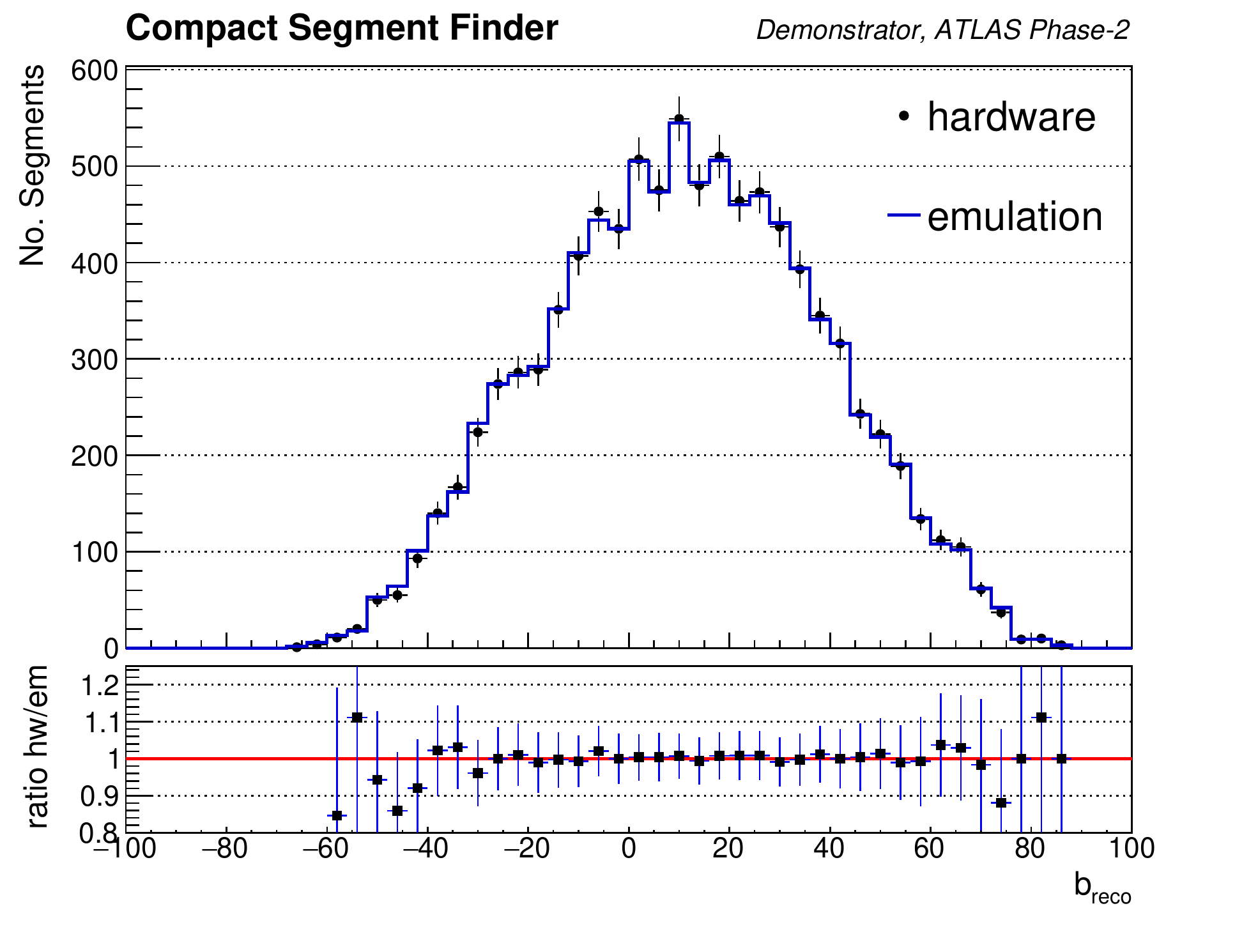}\label{<figure2>}}
     \caption{Distribution of segment candidate as a function of the fitted segment parameters ($\alpha, b$). Results obtained with the hardware demonstrator (dots) are compared with expectations from software emulation (blue line). A compatibility of 99.7\% has been observed, with a perfect matching of about 99.5\%. }
     \label{fw_em_param}
\end{figure}

Figure \ref{fw_em_param} shows the distributions of the reconstructed segment parameters in both hardware and emulation. The hardware is able to reconstruct about 99.7\% of the segments found by the emulation and in about 99.5\% of the cases, hardware and emulation create the same segment, with identical angular and spatial parameters.

\subsubsection{Segment Finding Efficiency}
\begin{figure}[hbpt]
\centering
	\subfloat[][]{\includegraphics[width=8cm]{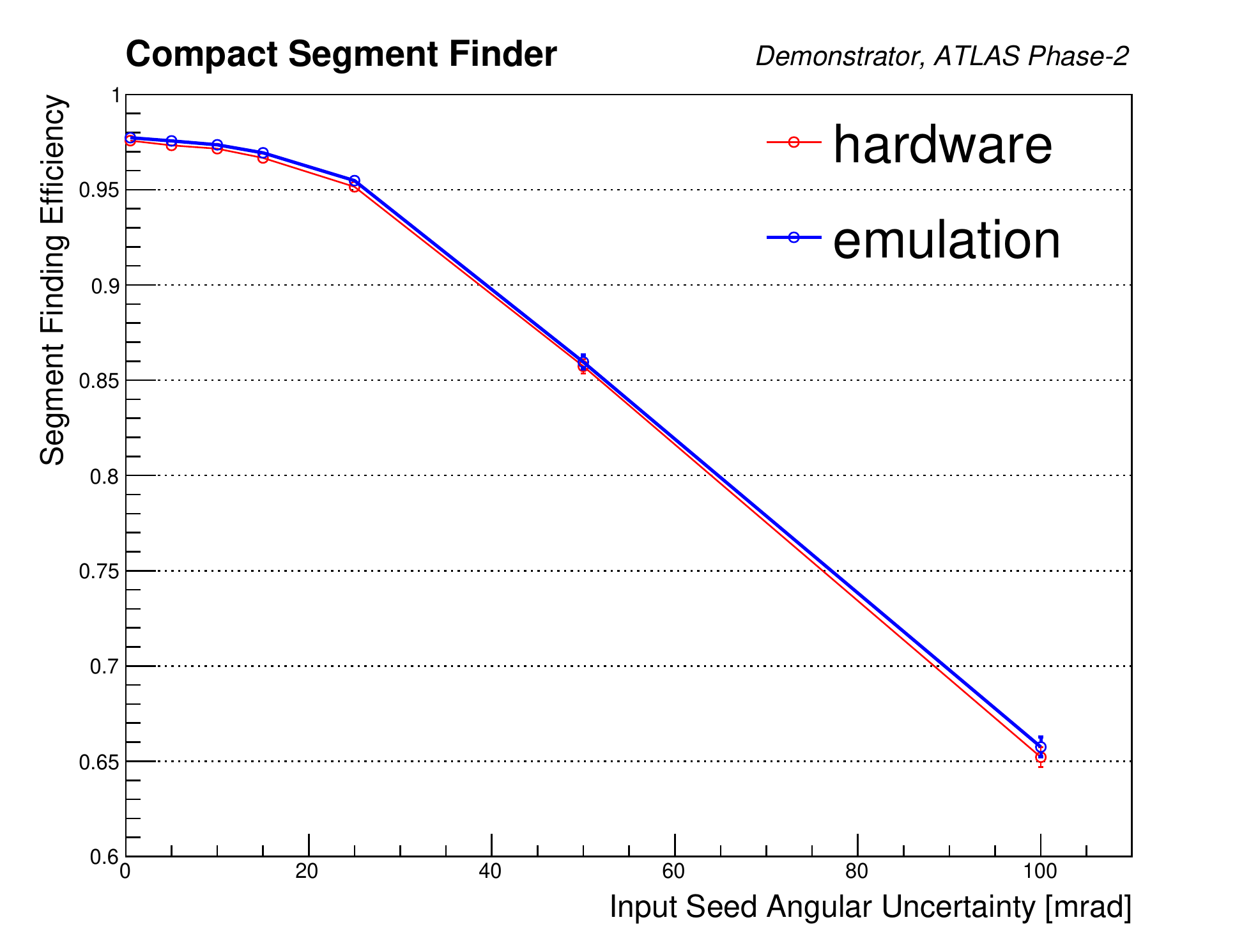}\label{<figure1>}}
     \subfloat[][]{\includegraphics[width=8cm]{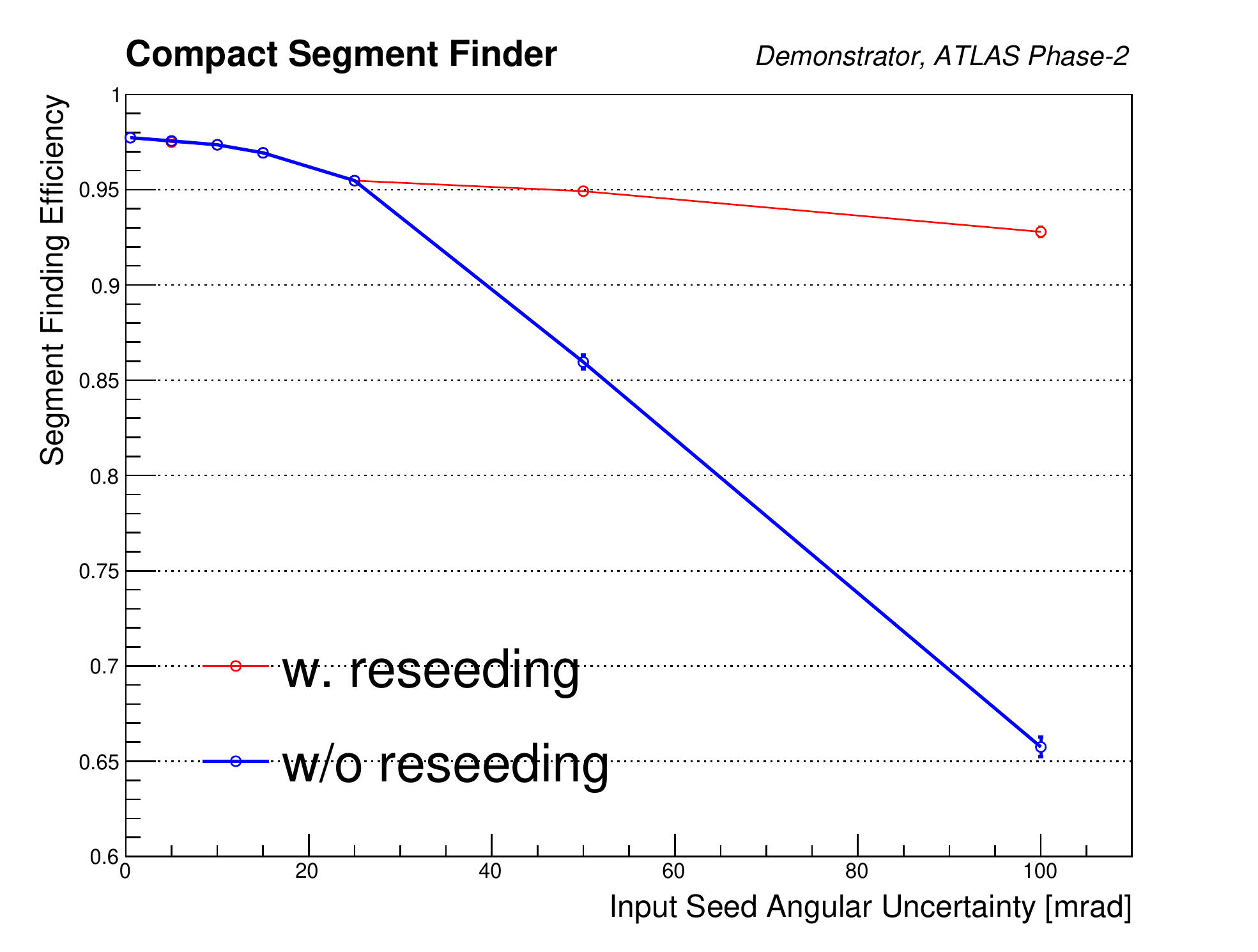}\label{<figure2>}}
\caption{Segment Finding efficiency as a function of the input seed accuracy. (a) Results obtained with the hardware demonstrator (red) are compared with expectations from software emulation (blue line). A good agreement is observed. (b) Simulated results obtained using a reseeding procedure in case of bad input hits. For reference results obtained without the reseeding are also shown. The reseeding procedure significantly improves the segment finding performance.}
\label{efficiency}
\end{figure}

The segment finding efficiency is measured with respect to all generated muon particles that produce at least two hits in each MDT multi-layer. According to the ATLAS Phase-II requirements \cite{ATLAS_TDAQ}, a muon segment is defined as correctly reconstructed if,

\begin{equation}
|\alpha_\text{reco}-\alpha_\text{gen}| < 3\,\text{mrad} \qquad \text{and}\qquad |b_\text{reco}-b_\text{gen}|<1\,\text{mm}\,,
\end{equation}
where $\alpha$ is the angle between the segment and the $z$-axis. Figure \ref{efficiency}a shows the segment finding efficiency as a function of the input seed accuracy, as measured in hardware and emulation. Both implementations produce similar results, although the hardware shows a slightly lower efficiency. This discrepancy is attributable to still present differences in the two implementations, which are foreseen to be fixed in future.

As expected the algorithm depends strongly on the input seed accuracy. A bad input seed with uncertainty larger than 25\,mrad can lead to a wrong filling of the HT array and, eventually, to a wrong selection of the hits for the final segment fitting. In any case, this situation is not realistic since the ATLAS RPCs are expected to provide an input seed with an angular resolution of about 20\,mrad in the worst case and the ATLAS TGCs have an even better resolution of the order of 5\,mrad.

Nevertheless, in case of a very bad seed or in the most pessimistic situation, where the pre-trigger is capable to provide only a RoI without seed parameters, a reseeding procedure can be adopted to retrieve this loss in efficiency. A simple way to recompute the input seed would be to use the average positions of the hits in the two MDT multi-layers and calculate the input slope by drawing a line between the two points. 

Figure \ref{efficiency}b shows the segment finding efficiency measured in emulation, with and without the reseeding. The reseeding procedure is able to compensate for the bad input seed accuracy, bringing the efficiency above 93\%. This simple reseeding procedure could be easily implemented on FPGA, with an estimated extra latency of twenty clock cycles (about 84\,ns at 240\,MHz).

\subsubsection{Segment Parameter Resolution}
\begin{figure}[hbpt]
\centering
	\subfloat[][]{\includegraphics[width=8cm]{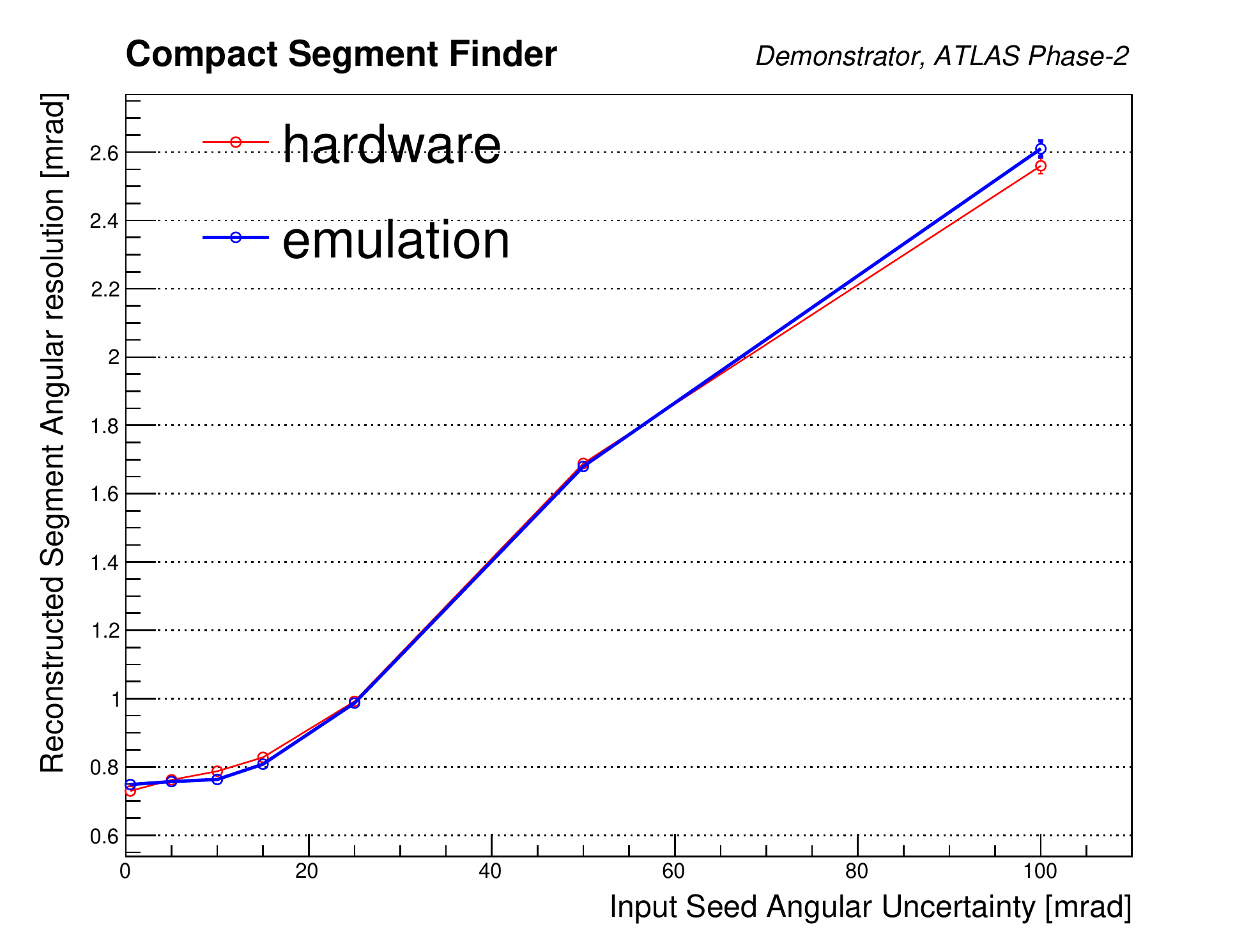}\label{<figure1>}}
     \subfloat[][]{\includegraphics[width=8cm]{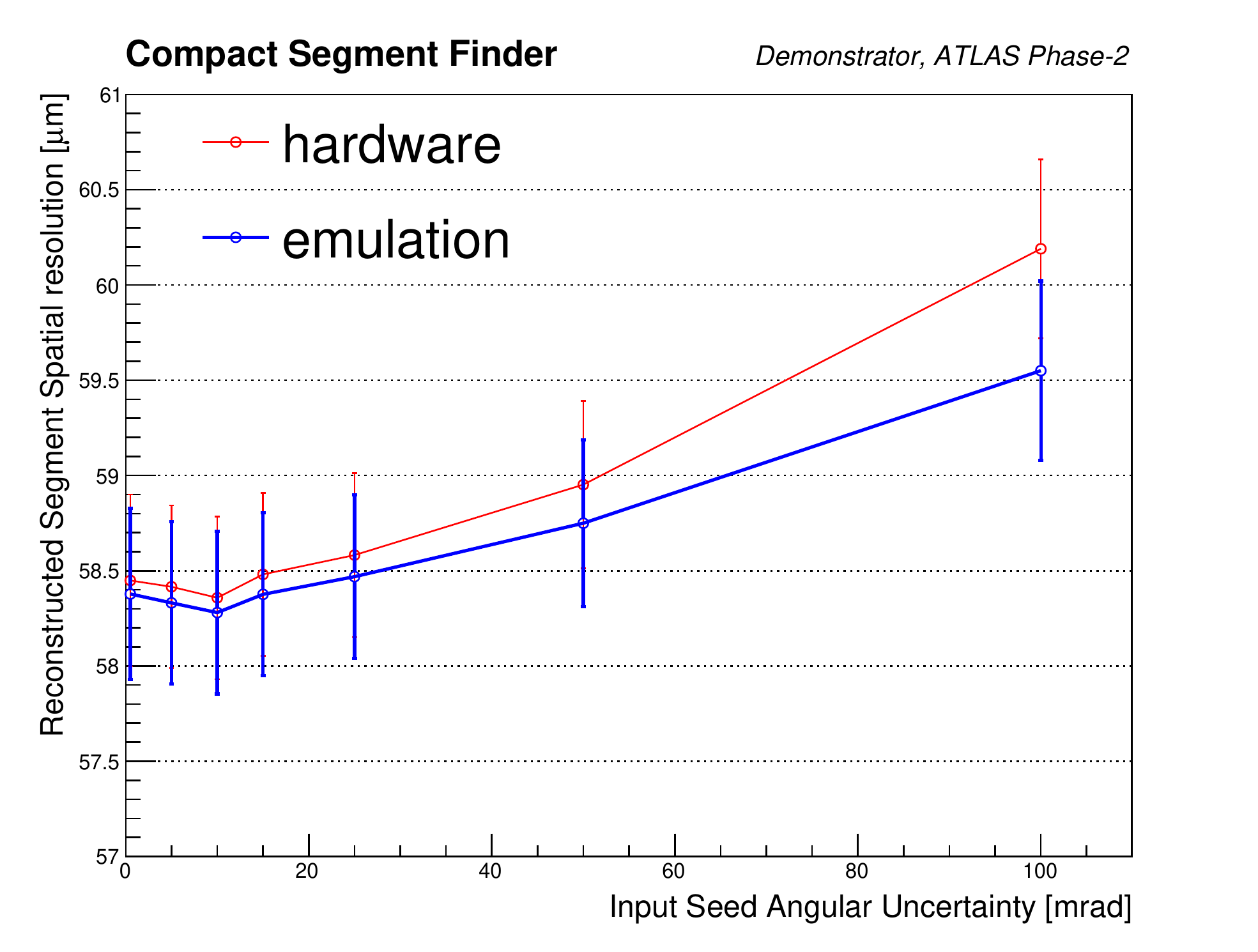}\label{<figure2>}}
\caption{Resolution of the angular (a) and spatial (b) segment parameters as a function of the input seed angular uncertainty. Results obtained with the hardware demonstrator (red) are compared with expectations from software emulation (blue). Similar performances are observed.}
\label{resolution}
\end{figure}

Figure \ref{resolution} shows the resolution of the two measured segment parameters ($\alpha, b$) for reconstructed segments for both hardware and emulation, as a function of the input seed angular uncertainty. The achieved resolutions are good enough to ensure an accurate measurement of the muon $p_T$. A reasonable level of agreement between hardware and emulation is observed, with remaining differences due to small discrepancies in some parts of the two implementations. A degradation in resolution for worse input seed is clearly visible, although it can be cancelled by using the   procedure previously illustrated.

\subsubsection{FPGA Resource Utilisation and Latency}
The implemented algorithm is very light in terms of resource utilisation. Table \ref{resources} shows the FPGA resource utilisation for each module of the firmware and for the entire Muon Track Finder Processor, as implemented on the Zynq 7045 SoC. The main utilised components are the DSPs, which are fundamental to run the segment fitting step. 

\begin{table}[hbpt]
\caption{Resource utilisation of the full Muon Track Finder processor and of the single firmware modules, as implemented in the Xilinx Zynq XC77045 SoC device. BRAM memory is used for the implementation of the DT Hit FIFOs. For comparison, also available logic resources of the Xilinx Zynq XC77045 and XC77100 are shown.}\label{resources}
\begin{center}
\begin{tabular}{ ccccc }
\hline 
& \textbf{LUTs} [$10^3$] & \textbf{DSPs} & \textbf{FFs} [$10^3$] & \textbf{BRAM (36 Kb)} \\
\hline
\textbf{Hit Processor} & 0 & 3 & 0 & 0 \\
\textbf{Histo A} & 1.2 & 0 & 0.5 & 0 \\
\textbf{Histo B} & 4.4 & 0 & 1.7 & 0 \\
\textbf{Fitter} & 2.6 & 59 & 2.4 & 0 \\
\textbf{$p_T$ calculator} & 1.5 & 3 & 0.1 & 17 \\
\hline
\textbf{Segment Finder} & 23.2 & 255 & 14.8 & 3.5 \\
\textbf{Muon Track Finder} & 71.7 & 837 & 44.5 & 27.5 \\
\hline
\textbf{Zynq 7045} & 218 & 900 & 437 & 545 \\
\textbf{Zynq 7100} & 277 & 2020 & 555 & 755 \\
\hline
\end{tabular}
\end{center}
\end{table}

The adopted firmware has also a low fixed latency, reconstructing one segment any 300\,ns with a running frequency of 240\,MHz. This relatively low latency provides enough time for the subsequent data transmission to the ARM processor and for the final $p_T$ computation, which requires approximately 360\,ns.

An alternative would be to implement also the transverse momentum computation on the PL. A $p_T$ calculator based on the sagitta method has been designed and implemented in the same Zynq 7045 SoC device. The implemented module has a fully pipelined design and is very lightweight in terms of resource usage (see Table \ref{resources}). The main utilised components are the DSPs for the equations and the BRAM, which stores the needed constants. The firmware has also a low-latency, requiring only seven clock cycles (ca. 30\,ns at 240\,MHz) to determine the muon transverse momentum. 

\section{Future Perspectives}
Although a system based on Zynq 7045 devices could be able to perform successfully the muon track finding task in a single trigger sector, it would be beneficial to adopt a slightly bigger chip. A better choice is given by the Xilinx Zynq XC7Z100, which has more than twice the number of DSP slices than the 7045. That would allow the implementation of the other three segment finder modules and, therefore, the reconstruction of another muon track. 

Further studies and improvements are also foreseen regarding the muon transverse momentum computation step, with a final decision to be taken on which implementation (FPGA or ARM) to adopt.

A prototype board based on Zynq/Kintex Ultrascale devices is under development within the ATLAS Phase-II MDT Trigger project, and it will be used to test the feasibility and the performance of a dimuon track finder processor. 

\section{Conclusions}
A compact algorithm for muon track reconstruction at future hadron collider experiments has been presented. The algorithm has been successfully tested by means of software emulation and a hardware demonstration. The algorithm is based on a 1D Hough Transform method to perform the pattern recognition and on a Simple Linear Regression technique for the segment fitting. 

The hardware demonstrator has successfully shown that it is possible to perform a full muon track reconstruction in a trigger sector composed of three muon chambers with good performance and low latency. The implemented algorithm is characterised by a low usage of FPGA resources and, therefore, it is an ideal candidate for hadron collider experiment purposes, where the large area of the system places stringent requirements also on the final costs.




\end{document}